# Gateway state-mediated, long-range tunnelling in molecular wires.[1,2]

Sara Sangtarash[a*‡], Andrea Vezzoli[b‡], Hatef Sadeghi[a], Nicolo' Ferri[b], Harry M. O'Brien[b], Iain Grace[a], Laurent Bouffier[b,3], Simon J. Higgins[b*], Richard J. Nichols[b], and Colin J. Lambert[a*]

[a] Quantum Technology Centre, Physics Department, Lancaster University, Lancaster LA1 4YB, UK

[b] Department of Chemistry, University of Liverpool, Crown Street, Liverpool L69 7ZD, UK

[‡] These authors contributed equally to this work

*Corresponding Authors:

E-mail:   shiggins@liverpool.ac.uk.

          s.sangtarash@lancaster.ac.uk

          c.lambert@lancaster.ac.uk

[1] Electronics Supplementary Information (ESI) available: synthetic procedures, single-molecule conductance measurements, additional theoretical details and calculations, $^1$H and $^{13}$C NMR Spectra.

[2] Data collected using EPSRC funding at Liverpool are archived at http://datacat.liverpool.ac.uk/id/eprint/198 or at DOI 10.17638/datacat.liverpool.ac.uk/198.

[3] Current address for Laurent Bouffier: Université de Bordeaux, ISM, CNRS, UMR 5255, Bordeaux INP, F-33400 Talence, France.

**Table of content (TOC) graphic:**

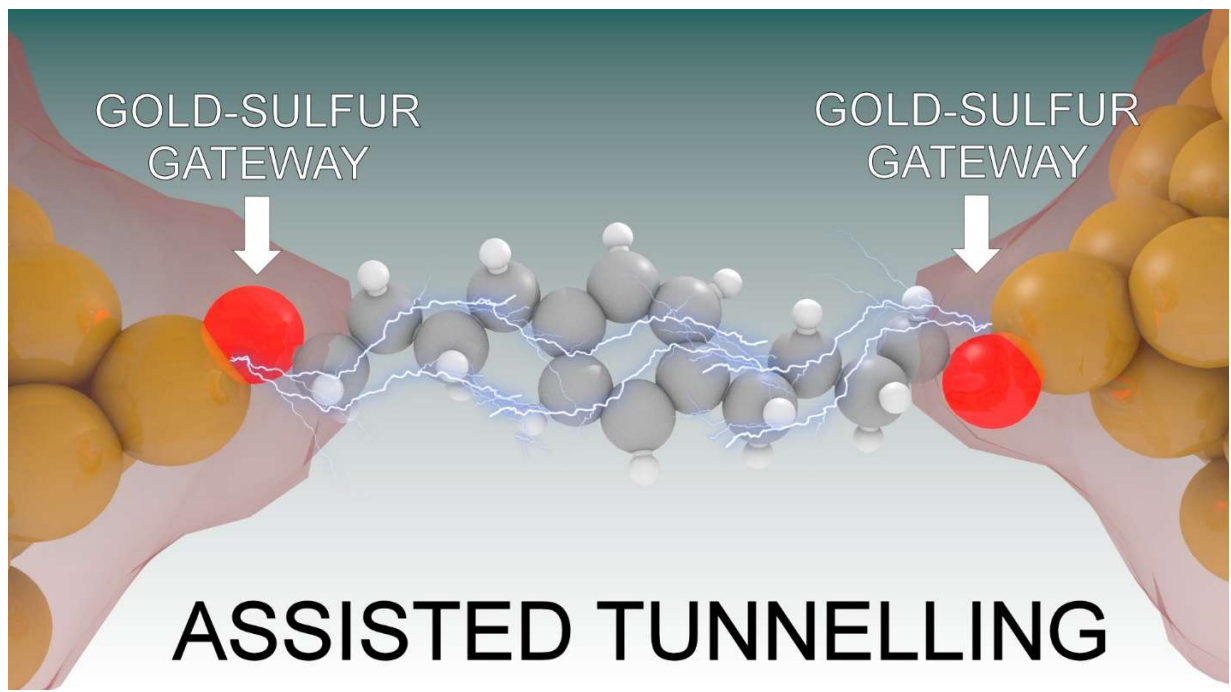

Gateway states' in Au / single-molecule / Au junctions profoundly attenuate the conductance decay with length for thiol-contacted alkyl-aromatic-alkyl systems

**Abstract**

If the factors controlling the decay in single-molecule electrical conductance $G$ with molecular length $L$ could be understood and controlled, then this would be a significant step forward in the design of high-conductance molecular wires. For a wide variety of molecules conducting by phase coherent tunneling, conductance $G$ decays with length following the relationship $G = Ae^{-\beta L}$. It is widely accepted that the attenuation coefficient $\beta$ is determined by the position of the Fermi energy of the electrodes relative to the energy of frontier orbitals of the molecular bridge, whereas the terminal anchor groups which bind to the molecule to the electrodes contribute to the pre-exponential factor $A$. We examine this premise for several series of molecules which contain a central conjugated moiety (phenyl, viologen or α-terthiophene) connected on either side to alkane chains of varying length, with each end terminated by thiol or thiomethyl anchor groups. In contrast with this expectation, we demonstrate both experimentally and theoretically that additional electronic states located on thiol anchor groups can significantly decrease the value of $\beta$, by giving rise to resonances close to $E_F$ through coupling to the bridge moiety. This interplay between the gateway states and their coupling to a central conjugated moiety in the molecular bridges creates a new design strategy for realising higher-transmission molecular wires by taking advantage of the electrode-molecule interface properties.

# Introduction

Understanding electron transport in metal−molecule−metal (MMM) junctions and identifying molecular wires whose conductance decays only slowly with length is important for the advancement of molecular electronics. The critical factors which determine conductance in a MMM junctions are the metal−molecule contacts and the structure of the molecular backbone.[1] While a wide variety of molecular backbones can be synthesised, the nature of the anchor groups that act as connectors to the metallic leads is limited by the strength of their interaction with the metal. As gold is the most widely used electrode material in molecular electronics, the choice of anchor can be made from moieties that can form X-Au covalent bonds, such as thiols[2,3] and carbodithioates,[4] moieties that react to give a C-Au bond, such as organostannanes[5] or diazonium salts,[6] and moieties that interact with gold with a coordination bond, such thiomethyls,[7–9] amines,[7,10] pyridines,[11–13] and phosphines[10].

Tunnelling theory predicts that conductance across a nanojunction should decay exponentially with its length, following a relationship $G = Ae^{-\beta L}$, where $L$ is the junction length and $A$ is a pre-exponential factor dependent on junction contacts and nature of metallic leads. The nature of the molecular wire bridging the two metallic leads has a strong effect on the exponential attenuation factor $\beta$, as demonstrated by Wold *et al.* in 2002.[14] Conjugated molecular wires such as oligophenylene exhibit conductance values that decay with increasing number of phenyl units to the extent of $\beta = 0.41$ Å$^{-1}$, and other conjugated systems such as oligophenyleneimine[15] and oligonaphthalenefluoreneimine[16] showed lower attenuation factors of 0.3 Å$^{-1}$ and 0.25 Å$^{-1}$, respectively. Extremely low values of $\beta$ were found in systems such as meso-to-meso bridged oligoporphyrins[13,17,18] (0.040 ± 0.006 Å$^{-1}$), axially-bridged oligoporphyrins[19] (0.015 ± 0.006 Å$^{-1}$), oligoynes[20] (0.06 ± 0.03 Å$^{-1}$), carbodithioate-capped oligophenylene-ethynylene[4] (0.05 ± 0.01 Å$^{-1}$), and extended viologens[21] (0.006 ± 0.004 Å$^{-1}$). Oligothiophenes, on the other hand, showed a more complex behaviour, with unusual conductance decay with the number of thiophene rings[22–24] and, in the case of longer oligothiophenes with alkylthiol linkers, water-dependent conductance and conductance decay.[25] A hopping charge-transport mechanism could explain the low value of β in some of these systems, but it is generally believed that tunnelling is dominant in short molecular wires. Transition to tunnelling to hopping has been observed in various systems, at a critical length ranging from 5 nm (oligonaphtalenefluoreneimine)[16] to 8 nm (oligothiophene).[26]

The length-dependent conductance of alkanedithiols (as archetypal saturated molecular wires) has been the subject of investigation by several research groups. Li *et al.*[27] reported exponential decrease of the conductance with molecular length with $\beta \approx 0.84$ Å$^{-1}$ for $N_{(CH2)} < 7$. Other studies with longer alkanedithiols showed that the conductance decay is less pronounced for shorter molecules ($N_{(CH2)} < 8$), whereas conductance decay is more rapid for longer lengths ($N_{(CH2)} > 8$).[28,29] Another study reports experimental decay constants $\beta \approx 0.94 - 0.96$ Å$^{-1}$.[30] Inclusion of heteroatom in the aliphatic alkyl chain to give oligoethers or oligothioethers resulted in negligible effect on $\beta$, with reported values of 1.11 (per atom unit) for alkanedithiols, 1.19 for oligoethers and 1.17 for oligothioethers.[31]

The above comprehensive experiments, combined with detailed modelling and material-specific transport calculations, take into account complex features introduced by metal-molecule contact,[32–34] orbital resonances, and other quantum mechanical effects that can strongly affect molecular conductance.[14,23,35–40] They have improved our understanding of the conductance decay with length in MMM junctions, but the effect of the molecular wire structure on the value of $\beta$ is still not completely understood. An important feature in the transport characteristics of alkanedithiols is the presence of a broad resonance, called in previous studies a "gateway state"[41] or "contact-level",[30] close in energy to the Fermi level of the metallic leads. In a systematic study of alkane molecular junctions with gold electrodes, Kim et al, reported a small peak close to the Fermi energy and a broad one about -1 eV from the Fermi energy, and they showed that the resonances are due to molecular orbitals localized on sulphur at these energies.[42] This peak is also present in the calculations of Hüser et al in the case of thiol end-groups connected to a single gold tip atom.[43]

We found that the presence of a central group attached via thiol-terminated alkane linkers to Au electrodes will magnify the effect of the resonance peak close to the Fermi energy, and we attribute this feature to atomic wave functions localised on sulfur atoms bound to the leads. The phenomenon is not limited to thiol contacts, and it has also been observed in MMM junctions with covalent, highly conducting C-Au contacts.[44] In what follows, we reveal the peculiar effect of these gateway orbitals on the decay constant $\beta$. In the conductance-length relationship $G = Ae^{-\beta L}$, the attenuation coefficient $\beta(E_F)$ is a property of the backbone and the value of the electrode Fermi energy $E_F$ relative to the frontier orbitals of the molecule, which determines the tunnelling gap for electrons passing from one electrode to the other. On the other hand, for a given $E_F$, it is often assumed that the coupling between the anchor groups and electrodes contributes to the prefactor *A* only. This assumption is surely correct in the asymptotic limit of large *L*, providing transport takes place by phase-coherent tunnelling. However, $\beta$ is usually obtained experimentally from the slope of plots of *ln(G)* versus *L*, for limited values of *L*, and the question of whether these values are sufficiently large is usually not addressed. However, there are some exceptions. For instance, Xie *et al*. have shown that, for junctions formed using conducting atomic force microscopy (C-AFM) measurements on monolayers of short oligophenyl molecules, the value of $\beta$ depended upon whether mono- or dithiols were employed,[45] being smaller in the latter case. In what follows, we refer to the slopes of such graphs as pseudo-attenuation coefficients and denote them $\beta'$. The main reason for doing this is that here we add methylene groups at either side of the central moiety, rather than having a homologous series incrementing by monomer units, the latter being the most widely used for the determination of attenuation factors $\beta$. The assumption that the coupling between the anchor groups and electrodes contributes to the prefactor *A* only has restricted the range of proposed strategies for manipulating $\beta$ to those which mainly rely on tuning $E_F$ by electrochemical or electrostatic gating, or by doping the backbone with electron donors or acceptors. The aim of the present paper is to demonstrate the counterintuitive result that the coupling between anchor groups and electrodes can contribute to both the $\beta'$ and the prefactor *A*. In this work we demonstrate this dependence, both experimentally and theoretically, by studying the length-dependent conductance of molecules containing a central conjugated moiety connected on either side to alkane chains of varying length.

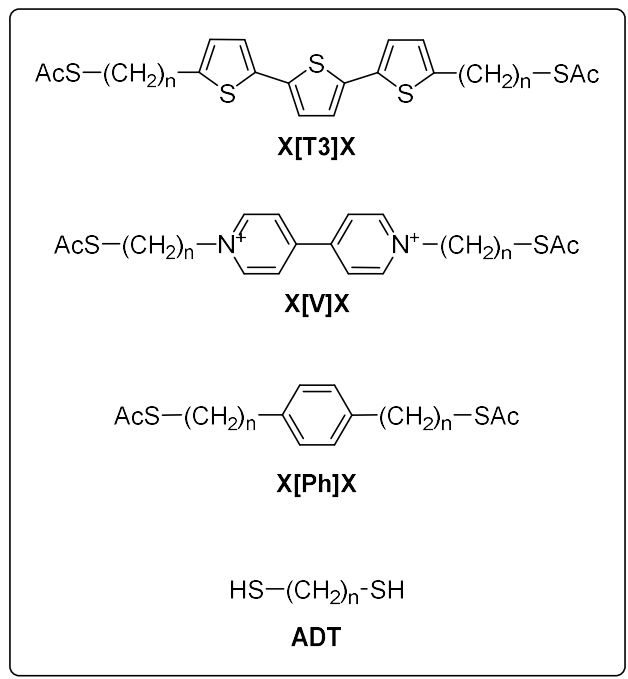

**Figure 1:** Structures and labelling of molecular systems discussed in this paper. X is the varying sidechain length (nCH$_2$) and the nature of the central conjugated unit is abbreviated between brackets.

As shown in Figure 1, the central units are chosen to be either an α-terthiophene (**X[T3]X**), a phenyl (**X[Ph]X**) or a viologen (**X[V]X**) moiety, and the alkane chains (varying in length from 1 to 9 methylene units each) are sulfur-capped to provide a strong connection to the gold electrodes. The nature of the central unit has been demonstrated to have a strong effect on molecular conductance,[46,47] and therefore we chose these three different moieties on the basis of their extent of conjugation and electron density, going from a poorly-conjugated, electron-deficient moiety such as a viologen salt (with a break in conjugation due to inter-ring torsion in its dication state[48]) to the well-conjugated, electron-rich α-terthiophene. Intuitively, one might expect the value of $β'$ for these molecular wires to approach the value determined for alkanedithiols ($β ≈ 1$ Å$^{-1}$). Surprisingly, in what follows we shall demonstrate that the presence of a conjugated moiety in the alkyl tunnelling barrier strongly affects $β'$, due to transport through "gateway states" and "coupling states", the magnitude of which depends on the nature of the conjugated system.

## Results and Discussion

The series of molecular wires shown in Figure 1 were synthesised and characterised using common synthetic laboratory techniques (see ESI for synthetic procedures). The STM-based $I(z)$ technique[49] (details in the Methods section) was used to measure the conductance of the molecular wires presented in this work, and the more widely used STM-BJ technique[11] was used as comparison for the most conductive molecular wire (more information in the ESI). In brief, a gold tip is moved towards a gold surface with a sub-monolayer of the target molecule and then retracted to yield current ($I$) – distance ($z$) traces that show a number of features characteristic of MMM junctions, such as steps and plateaus. Hundreds of such conductance-distance traces are collected and subsequently compiled in histograms bearing a distribution of conductance values. Peaks in the histograms were fitted to a Gaussian distribution to determine the most probable conductance, expressed in nS. Data for the **X[Ph]X** class of molecular wires was taken after Brooke *et al.*,[41] and data for the alkanedithiol series was taken from Haiss *et al.*[29] The experimentally determined conductance values were then plotted as *ln(G) vs* length, and a linear fitting was used to obtain the *β'* attenuation and its standard deviation is used as error.

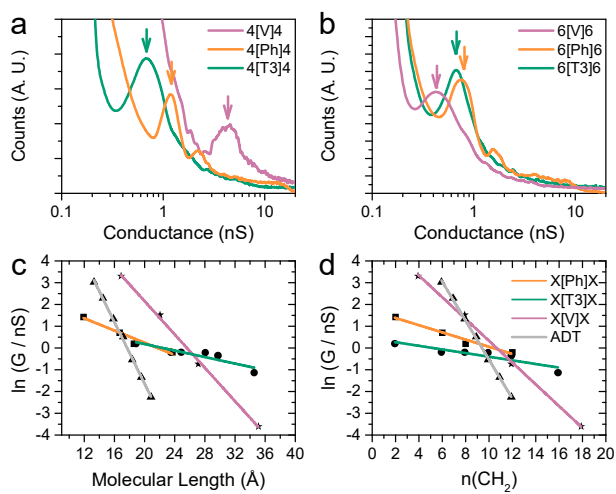

**Figure 2:** Examples of conductance histograms for (a) **4[X]4** and (b) **6[X]6**. Data is displayed on a logarithmic scale, bin size 0.05 nS. Histograms are normalised to the total number of *I(z)* scans selected (791 for **4[V]4**, 505 for **4[Ph]4**, 592 for **4[T3]4**, 640 for **6[V]6**, 946 for **6[Ph]6** and 748 for **6[T3]6**). ln(G) vs. length plots and linear fitting of data with (c) molecular length expressed in Å and (d) in number of methylene units in the sidechains. Molecular length is calculated as distance between two gold atoms tethered to the sulfur ends of the molecules in their lowest energy (all trans) conformation, using Wavefun Spartan® '14. **4[Ph]4** and **6[Ph]6** show a smaller peak at higher conductance which is due to multiple molecular bridging in the Au-Au gap. All data recorded at 300 mV tip-substrate bias. See Figures S1-S10 (ESI) for individual histograms.

Example of results of single-molecule conductance measurements for the molecular wires capped with protected thiol functions are presented in Figure 2a and Figure 2b (see ESI for further data), and the observed $\beta'$ decay is summarised in Figure 2c and Figure 2d. The **X[T3]X** series showed a very shallow conductance decay of $0.06 \pm 0.01$ Å$^{-1}$ ($0.07 \pm 0.02$ per methylene unit), more than one order of magnitude smaller than the value found in alkanedithiols. Measurements on **X[Ph]X** and **X[V]X** series gave a higher $\beta'$ value of $0.14 \pm 0.02$ Å$^{-1}$ ($0.17 \pm 0.03$ per methylene unit) and $0.39 \pm 0.01$ Å$^{-1}$ ($0.52 \pm 0.01$ per methylene unit), respectively. The **X[V]X** system has already been the subject of theoretical[50] and experimental[51] studies, and the published $\beta'$ value is slightly higher, at $0.59 - 0.61$ Å$^{-1}$ ($0.66 - 0.76$ per methylene unit). It must be noted, however, that these published values were obtained using the limited length interval from **5[V]5** to **8[V]8**, while the data presented in this work spans a significantly larger interval (**2[V]2** to **9[V]9**). Previous results obtained from the **X[V]X** system[47,52] have been interpreted with the Kuznetsov-Ulstrup model,[53] a sequential two-channel mechanism due to non-resonant charge transport under electrochemical control. This model only applies when the system is "electrochemically tuned" such that molecular redox levels are in close energetic proximity to the contact Fermi levels. Under the open circuit conditions of the two-terminal measurements presented here, this is not the case and instead phase coherent tunnelling best describes the transport.

In order to explain the experimental data, we performed density functional theory - non-equilibrium Green's function (DFT-NEGF) theoretical calculations to model charge transport across the MMM junctions. To calculate the conductance of the junction consisting of two gold electrodes connected to the molecule, the optimal geometry and ground state Hamiltonian were obtained using SIESTA[54] implementation of density functional theory and the room-temperature electrical conductance was calculated using the GOLLUM[55] code (more details in the Methods section). The computed conductance $G$ depends on the Fermi energy $E_F$ of the contacts, and since the value of $E_F$ relative to the frontier orbital energies of the molecule are not necessarily accurately predicted by DFT, in what follows we present results for $G$ over a range of values of $E_F$, centred on the DFT-predicted value of $E_F = 0$.

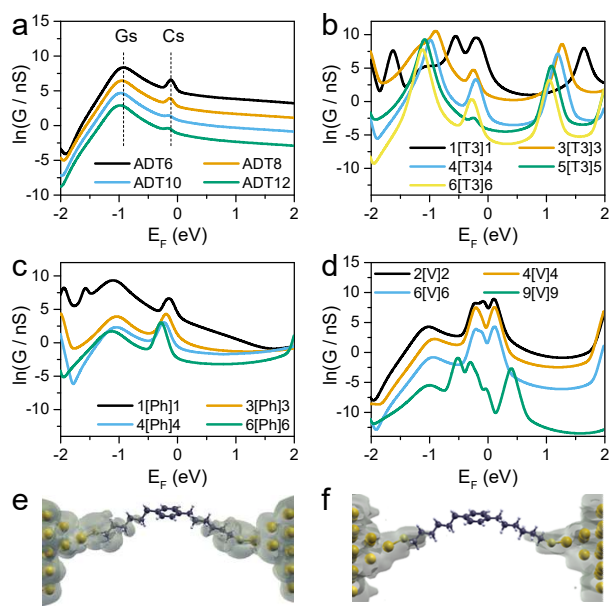

**Figure 3:** Calculated room-temperature conductance of (a) **X[Ph]X**, (b) **X[T3]X**, (c) **ADT**, and (d) **X[V]X** vs. Fermi energy for alkyl chains with different numbers of CH$_2$ groups. $E_F$ = 0 corresponds to the DFT-predicted Fermi energy. LDOS at -0.15 eV (e) showing Cs, and at -1.11 eV (f) showing Gs for **6[Ph]6**.

Figure 3 shows the calculated conductance of the series of molecular wires, **ADT** for n(CH$_2$) = 6, 8, 10 and 12 (Figure 3a), **X[T3]X** where X = 1, 3, 4, 5 and 6 (Figure 3b), **X[Ph]X** where X = 1, 3, 4 and 6 (Figure 3c), and **X[V]X** where X = 2, 4, 6 and 9 (Figure 3d). As shown in the conductance vs. Fermi energy plots for alkane dithiols (Figure 3a), a broad resonance is present in the HOMO-LUMO gap, at about $E_F$ = -1 eV (labelled "Gs" in Figure 3a). This feature has been previously assigned to "gateway" or "Au-S" states located on the sulfur anchors either by theoretical[30,37,43] or spectroscopic means.[56] Additional smaller but sharper resonances arise in our calculations very near to $E_F$ = 0 (labelled "Cs" in figure 3a), and we attribute these to the strong coupling between the two gateway states, through the molecular backbone ("coupling states"; see local density of states plots in Figure 3e and 3f). These two features have already been discussed in the literature,[42,43] and are present in all the calculations performed on the compounds presented in this study with thiol contacts, but their combined magnitude is more pronounced in the presence of central moieties comprising a α-terthiophene (Figure 3b) or a phenyl ring (Figure 3c) and, as shown in Figure 4, they lead to a low *β'* value for these molecules.

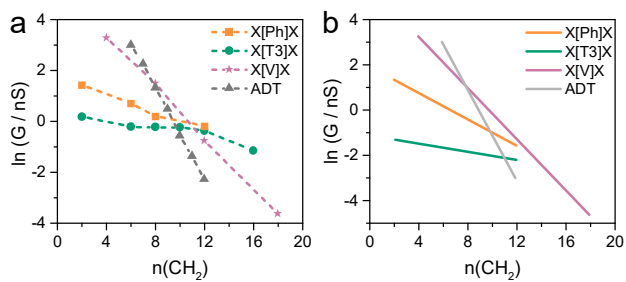

**Figure 4**: Comparison of experimental (a) and calculated (b) natural logarithm of conductance versus the number of methylene units in the side chains for the four series of molecular wires considered in this study. See Figures S1-S10 (ESI) for individual histograms.

Figure 4b shows results for conductance values ($G$) obtained using the DFT-predicted Fermi energy and the slopes of these plots of $ln(G)$ versus number of methylene units yields $\beta'$ for each molecular series. For direct comparison, the dash lines in Figure 4a correspond to the experimental results. The values of the theoretical attenuation factors are: $\beta'_{X[T3]X} = 0.086$, $\beta'_{X[Ph]X} = 0.20$, $\beta'_{X[V]X} = 0.56$ and $\beta_{ADT} = 0.93$ per methylene unit. It is then clear that the trends of the calculated $\beta$ and $\beta'$ are in good agreement with the measured one. As shown in Figure 4, in both theory and experiment the value of $\beta_{ADT} > \beta'_{X[V]X} > \beta'_{X[Ph]X} > \beta'_{X[T3]X}$.

Since theory predicts that the low $\beta'$ values in **X[Ph]X** and **X[T3]X** are due to the transport through gateway/coupling states located on the sulfur atoms, we expect to dramatically increase the conductance attenuation if these states are removed. Thiomethyl (-SMe) groups interact with gold via a coordination bond between the metal orbitals and the lone pair localised on the sulfur atom, and the absence of a strong covalent bond should give no gateway or coupling states in the transmission curves. The calculated room-temperature conductance as function of energy is presented for **X[Ph]X** with thiol anchors in Figure 5a and for **X[Ph]X-SMe** (with methyl thioether anchors on both sides) in Figure 5b. As expected, in the presence of -SMe anchors there are no additional resonances in the HOMO-LUMO gap, and this results in a greatly increased value of $\beta'$.

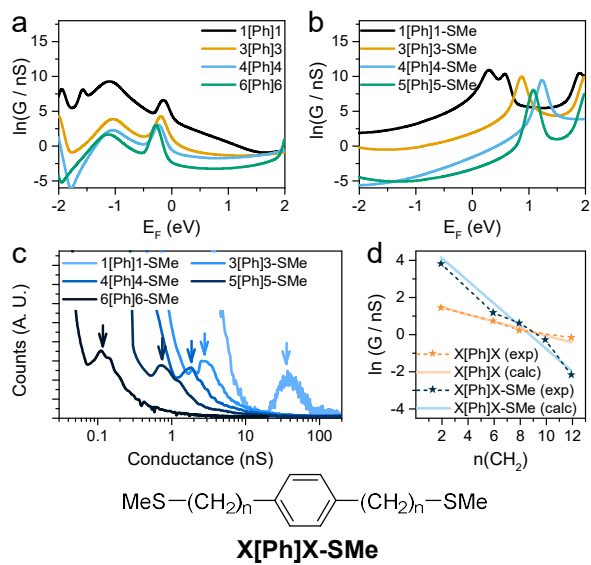

**Figure 5:** Calculated conductance of (a) Au-**X[Ph]X**-Au and (b) Au-**X[Ph]X-SMe**-Au at room temperature with predicted DFT-gap from Kohn-Sham mean field Hamiltonian. (c) Experimental **X[Ph]X-SMe** conductance histograms. Data is displayed on a logarithmic scale, bin size 0.01 nS. Histograms are normalised to the total number of *I(z)* scans selected (607 for **1[Ph]1-SMe**, 656 for **3[Ph]3-SMe**, 501 for **4[Ph]4-SMe,** 566 for **5[Ph]5-SMe** and 513 for **6[Ph]6-SMe**). (d) Experimental and calculated conductance for **X[Ph]X** (orange) and **X[Ph]X-SMe** (blue) vs. number of CH$_2$ units. Structure of the **X[Ph]X-SMe** system is shown for clarity. See Figures S11-S15 (ESI) for individual histograms.

To confirm the theoretical findings, we synthesised the series of molecular wires bearing a phenyl central unit and alkyl spacers of varying length with thiomethyl contacts (**X[Ph]X-SMe**), and measured their conductance (Figure 5c). The results confirmed the theoretical prediction, with an increased attenuation factor $\beta' = 0.50 \pm 0.04$ Å$^{-1}$ (0.56 ± 0.05 per methylene unit) upon removal of the gateway/coupling states (Figure 5d). Thiomethyl is not the only contact group that increases the $\beta'$ value in these dialkyl benzene compounds, and an even higher value of > 1 per methylene unit has been reported, for instance, in carboxylic acid-capped molecular wires.[57]

The role of these additional states can be further described from an analytical perspective, by using a simple theory which captures their effect in terms of two dimensionless parameters. In the low-voltage and low-temperature limit, the electrical conductance of a single molecule is given by

$$G = G_0 T(E_F) \qquad (1)$$

where $G_0 = \left(\frac{2e^2}{h}\right)$ is the quantum of conductance, $E_F$ is the Fermi energy of the electrodes and $T(E_F)$ is the transmission coefficient for electrons of energy $E_F$ passing from one electrode to the other, via the molecule. The transmission coefficient $T(E_F)$ is a property of the whole system comprising the leads, the molecule and the contact between the leads and the molecule. For example for the model structure sketched in Figure 6a, in which electrons of energy $E_F$ pass through a single molecular orbital of energy $E_0$, $T(E_F)$ is given by the Breit-Wigner formula[1]

$$T(E_F) = \frac{B}{(\varepsilon_F^2 + 1)} \qquad (2)$$

where $B = 4\Gamma_1\Gamma_2/(\Gamma_1+\Gamma_2)^2$ and $\varepsilon_F = [E_F - E_0 - (\sigma_1 + \sigma_2)]/[\Gamma_1 + \Gamma_2]$. In this expression, $E_0$ is the energy of the molecular orbital when the molecule is isolated from the electrodes, while $\sigma_1$ and $\sigma_2$ describe the shift in the resonance energy due to contact with the left (1) and right (2) electrodes, respectively. Equation (2) reveals that $T(E_F)$ is a maximum when $E_F$ satisfies the on-resonance condition $E_F = E_0 + (\sigma_1 + \sigma_2)$. When plotted as a function of $E_F$, the half width at half maximum of $T(E_F)$ is $\Gamma_1 + \Gamma_2$, where $\Gamma_1$ and $\Gamma_2$ describe the contributions to this broadening due to contact with the left and right electrodes. Equation (2) is valid provided the energy-level spacing of the molecule is greater than $\Gamma_1 + \Gamma_2$. Although $T(E_F)$ depends on six material-specific parameters ($E_F$, $E_0$, $\sigma_1$, $\sigma_2$, $\Gamma_1$, $\Gamma_2$), which themselves depend on structural and chemical parameters describing the junction, equation (2) reveals that all generic features are captured by only two dimensionless parameters, $B$ and $\varepsilon_F$. In the case of a symmetric molecule attached symmetrically to identical leads, $\Gamma_1 = \Gamma_2 = \Gamma$, $\sigma_1 = \sigma_2 = \sigma$ and hence $B = 1$. Therefore on resonance (i.e. when $\varepsilon_F = 0$), $T(E_F) = 1$. On the other hand, in the case of an asymmetric junction, where for example $\Gamma_1 \gg \Gamma_2$, the on-resonance values of $T(E_F)$ is less than unity.

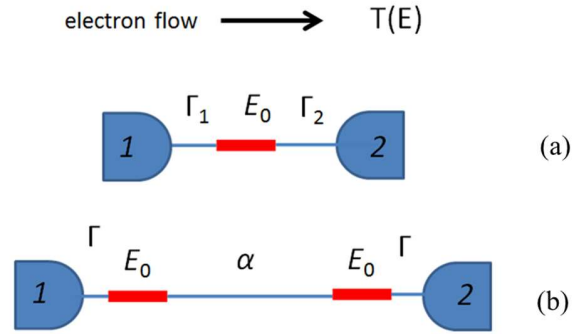

**Figure 6**: (a) A sketch of the model structure described by equation (2), comprising a single molecular orbital connected to electrodes 1 and 2. (b) Model structure with two degenerate coupling orbitals of energy $E_0$ connected to electrodes 1 and 2, and described by equation (3).

Now consider the case of a molecule containing two gateway orbitals of equal energy $E_0$, coupled to each other by a tunnel barrier such as an alkyl chain represented by a tunnelling matrix element $\alpha$, which decays exponentially with the length of the chain (Figure 6b). Each of the orbitals is connected separately to electrodes 1 and 2. Since the energy spacing between these orbitals is zero (and therefore less than $\Gamma_1 + \Gamma_2$), equation (1) cannot be used and the mathematical description of $T(E_F)$ is more complex (equation S1 in the ESI).[58] For simplicity, we consider only the case of a symmetric junction, for which the formula of equation S1 reduces to

$$T(E_F) = A f(\varepsilon_F, a) \qquad (3)$$

where $\varepsilon_F = (E_F - E_0 - \sigma)/\Gamma$, $a = |\alpha|/\Gamma$, $A = 4a^2$ and $f(\varepsilon_F, a) = \frac{1}{[(\varepsilon_F - a)^2 + 1][(\varepsilon_F + a)^2 + 1]}$.

Like equation (2), this expression also involves two dimensionless parameters, namely $a$ and $\varepsilon_F$. However, as shown in Figure 7, unlike equation (2), when plotted against $\varepsilon_F$, equation (3) possesses two maxima when $|a| > 1$. These maxima occur at $\varepsilon_F = \pm(a^2 - 1)^{1/2}$, which are associated with bonding and anti-bonding combinations of the two gateway orbitals, induced by the coupling $\alpha$. Since $\alpha$ decreases with the length of the alkyl bridge, two maxima (the "coupling states") are present for short molecules (*i.e.* for larger $\alpha$) and merge into a single maximum for longer molecules. This splitting, for instance, also appears in DFT calculations of short **ADT** (Figure S24 of the ESI).[43]

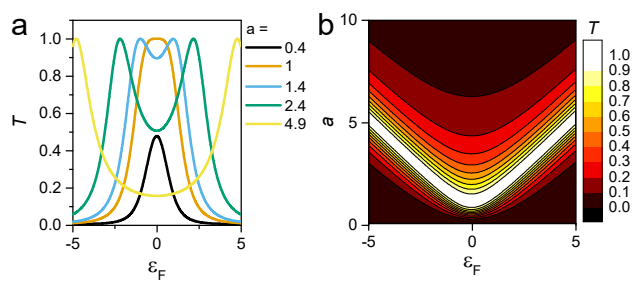

**Figure 7**: (a) Plots of equation (3) versus $\varepsilon_F$ for various values of $a$ and (b) contour diagram of the right-hand side of equation (3), where the coupling parameters $a$ and $\varepsilon_F$ are plotted as a function of T (colour bar). The splitting is evident in the contour plot for $a > 1$.

Both the splitting and position of these maxima are sensitive to the length and conformation of the tunnelling bridge, and this length dependence is different for **X[Ph]X** compared with **ADT**, due to conformational changes induced by the presence of the phenyl ring in the latter. These overall variations of $a$, $\Gamma$, $\varepsilon_F$ and $T(E_F)$ with the number of alkyl units leads to a steeper slope for $\ln f(\varepsilon_F, a)$ in **X[Ph]X**, which cancels the trend in $\ln a$ and decreases in the value of $\beta'$, in agreement with the experimental results and DFT calculations (further details on the parameters in section 8 of the ESI).

## Conclusions

The above results demonstrate that when a conjugated central unit is sandwiched between two insulating alkyl chains the decay in conductance with the length of the chains is much shallower than that of alkyl chains alone. For example, the beta factor of an alkanedithiol is $\beta = 0.9$ Å$^{-1}$, whereas in the presence of a phenyl ring central unit, this decreases to $\beta' = 0.18$ Å$^{-1}$ and in the presence of an α-terthiophene central unit, it further decreases to $\beta' = 0.07$ Å$^{-1}$. DFT and NEGF transport calculations demonstrated that this shallow length dependence is linked to charge transport assisted by additional states localised near the Au-S contact. To further investigate this phenomenon, we replaced the covalent-bonding thiol anchor with a coordination-bonding thiomethyl, which does not possess additional states near the Fermi energy. The predicted increase to $\beta' = 0.50$ Å$^{-1}$ for thiomethyl-terminated dialkylbenzene (**X[Ph]X-SMe**) molecular wires was confirmed experimentally. To underpin this new concept of gateway-driven conductance attenuation, we also introduced a simple model involving two dimensionless parameters, whose length dependence encapsulates these trends.

## Methods

**Syntheses:** Molecular wires used in this study were synthesised using common synthetic laboratory techniques. The syntheses of **6[T3]6**[59] and the two longest viologens (**6[V]6** and **9[V]9**) are described elsewhere.[60] Synthetic procedures and characterisation data for previously unreported compounds are provided in the Electronics Supplementary Information.

**Conductance measurements:** The conductance of molecular junctions was determined using the STM *I(z)* technique as described previously in the literature.[49,61] An adsorbed layer of the target molecule is formed on a flame-annealed gold-on-glass substrate (Arrandee Metal GmbH, DE – 250 nm Au, 2.5 nm Cr, 0.7 mm borosilicate glass) by immersion in a dilute (10$^{-4}$ M) solution in CH$_2$Cl$_2$. The thioacetate moiety cleaves spontaneously in the presence of a Au substrate.[62] The substrate is rinsed with copious ethanol to remove physisorbed molecules and dried under a stream of argon. An STM (former Molecular Imaging PicoSPM I, now 4500 SPM, Keysight Technologies Inc., USA) is used to drive an Au tip (Goodfellow Cambridge Ltd., UK – 99.99+%, 0.25 mm) close to the gold substrate at a defined setpoint current and under constant bias, so that junctions can form, and rapidly retracted (40 nm s$^{-1}$), while a current ($I$) vs. distance ($z$) curve is recorded. The process is repeated many times, and hundreds of such junction making and breaking curves are analysed statistically in histograms to yield a distribution of conductance values. Spurious traces with no evidence of junction formation (plateaux and steps) were discarded to avoid ambiguity and reduce noise. The average hit rate (percentage of scans showing evidence of junction formation) is 10-15 %, depending on the molecular wire. Plateaux in current-distance curves result in peaks in the histogram, and a Gaussian fit was used to determine the most probable conductance value.

**Theoretical calculations:** The Hamiltonian of the structures described in this paper were obtained using density functional theory as described below or constructed from a simple tight-binding model with a single orbital per atom of site energy $\varepsilon_0 = 0$ and nearest neighbour couplings $\gamma = -1$.

**DFT calculations:** The optimized geometry and ground state Hamiltonian and overlap matrix elements of each structure was self-consistently obtained using the SIESTA[54] implementation of DFT. SIESTA employs norm-conserving pseudo-potentials to account for the core electrons and linear combinations of atomic orbitals to construct the valence states. The generalized gradient approximation (GGA) of the exchange and correlation functional is used with the Perdew-Burke-Ernzerhof parameterization (PBE)[63] a double-ζ polarized (DZP) basis set, a real-space grid defined with an equivalent energy cut-off of 250 Ry. The geometry optimization for each structure is performed to the forces smaller than 40 meV/Ang.

**Transport calculations:** The mean-field Hamiltonian obtained from the converged DFT calculation or a simple tight-binding Hamiltonian was combined with our implementation of the non-equilibrium Green's function method, the GOLLUM[55] code, to calculate the phase-coherent, elastic scattering properties of the each system consist of left (source) and right (drain) leads and the scattering region. The transmission coefficient T(E) for electrons of energy E (passing from the source to the drain) is calculated via the relation $T(E) = \text{Trace}(\Gamma_R(E)G^R(E)\Gamma_L(E)G^{R\dagger}(E))$. In this expression, $\Gamma_{L,R}(E) = i\left(\Sigma_{L,R}(E) - \Sigma_{L,R}^{\dagger}(E)\right)$ describe the level broadening due to the coupling between left (L) and right (R) electrodes and the central scattering region, $\Sigma_{L,R}(E)$ are the retarded self-energies associated with this coupling and $G^R = (ES - H - \Sigma_L - \Sigma_R)^{-1}$ is the retarded Green's function, where H is the Hamiltonian and S is the overlap matrix. Using the obtained transmission coefficient (T(E)), the conductance could be calculated by the Landauer formula ($G = G_0 \int dE\, T(E)(-\partial f/\partial E)$) where $G_0 = 2e^2/h$ is the conductance quantum, $f(E) = (1 + \exp((E - E_F)/k_B T))^{-1}$ is the Fermi-Dirac distribution function, T is the temperature and $k_B$ = 8.6 × 10$^{-5}$ eV/K is the Boltzmann's constant.

## Associated Content

**Electronics Supplementary Information:**

- Synthetic procedures, single-molecule conductance measurements, additional theoretical details and calculations, $^1$H and $^{13}$C NMR Spectra.

**Experimental Data:** Data collected using EPSRC funding at Liverpool are archived at http://datacat.liverpool.ac.uk/id/eprint/198 or at DOI 10.17638/datacat.liverpool.ac.uk/198.

## Author Information:

**Corresponding authors**

*E-mail: shiggins@liverpool.ac.uk

*E-mail: s.sangtarash@lancaster.ac.uk

*E-mail: c.lambert@lancaster.ac.uk

**Author Contributions**

S.J.H. and R.J.N. conceived the project. A.V., N.F., H.M.O. and L.B. synthesized the compounds and characterised them. A.V. and N.F. performed the STM measurements, and A.V. analysed the experimental data. S.S., H.S. and C.J.L. developed the theoretical explanation, and S.S. performed the calculations. S.J.H., C.J.L. and R.J.N. supervised the project. A.V. and S.S. wrote the manuscript. All authors discussed the results and commented on the manuscript.

**Notes**

The authors declare no competing financial interest.

## Acknowledgment

This work was supported by the UK EPSRC under grant EP/K001507/1 "Transition-edge sensors", EP/H035818/1 and EP/H035184/1 "Medium effects in single-molecule electronics", EP/J014753/1 "Ultra-high resolution, ultra-sensitive multifunctional ballistic nano sensors for the simultaneous detection of magnetic, electric and optical fields", and EP/M005046/1 "Single-molecule photo-spintronics". We also acknowledge support by the European Commission (EC) FP7 ITN "MOLESCO" network (project no. 606728).